\documentclass[prb,twocolumn,aps,showpacs,fixfloats]{revtex4}
\usepackage{graphicx}
\usepackage{bm}
\usepackage{amsmath,amssymb}
\usepackage{subfigure}
\usepackage{float}
\usepackage{latexsym}
\usepackage{epstopdf}
\usepackage{color}
\usepackage{enumerate}
\usepackage{pdfpages}

\newcommand{\s}{\scriptscriptstyle}

\begin{document}

\title {Smearing of the quantum  anomalous Hall effect due to statistical fluctuations of magnetic dopants  }

\author{Z. Yue and M. E. Raikh }

 \affiliation{Department of Physics and
Astronomy, University of Utah, Salt Lake City, UT 84112, USA
}

\begin{abstract}
Quantum anomalous Hall effect (QAH) is induced by
substitution of a certain portion, $x$,
of Bi atoms in a BiTe-based insulating parent compound
by magnetic ions (Cr or V).
We find the  density of in-gap  states, $N(E)$, emerging as a result of statistic fluctuations
of the composition, $x$, in the vicinity of the transition point, where the {\em average} gap,
${\overline E}_g$, passes through zero.
 Local gap follows the fluctuations of $x$. Using the instanton approach, we show that,
 near the gap edges, the tails
 are exponential, $\ln{N(E)} \propto - \big({\overline E}_g-|E|\big)$, and the tail states
 are due to small gap reduction. Our main finding is that, even when the smearing magnitude exceeds
 the gap-width, there exists are semi-hard gap around zero energy, where
 $\ln {N(E)} \propto -\frac{{\overline E}_g}{|E|}\ln \Big(\frac{{\overline E}_g}{|E|}\Big)$.
 The states responsible for $N(E)$ originate from local gap reversals within narrow rings.
 The consequence of semi-hard gap is the Arrhenius, rather than variable-range hopping, temperature
 dependence of the diagonal conductivity at low temperatures.
%

\end{abstract}
\maketitle

\begin{figure}
\includegraphics[width=84mm]{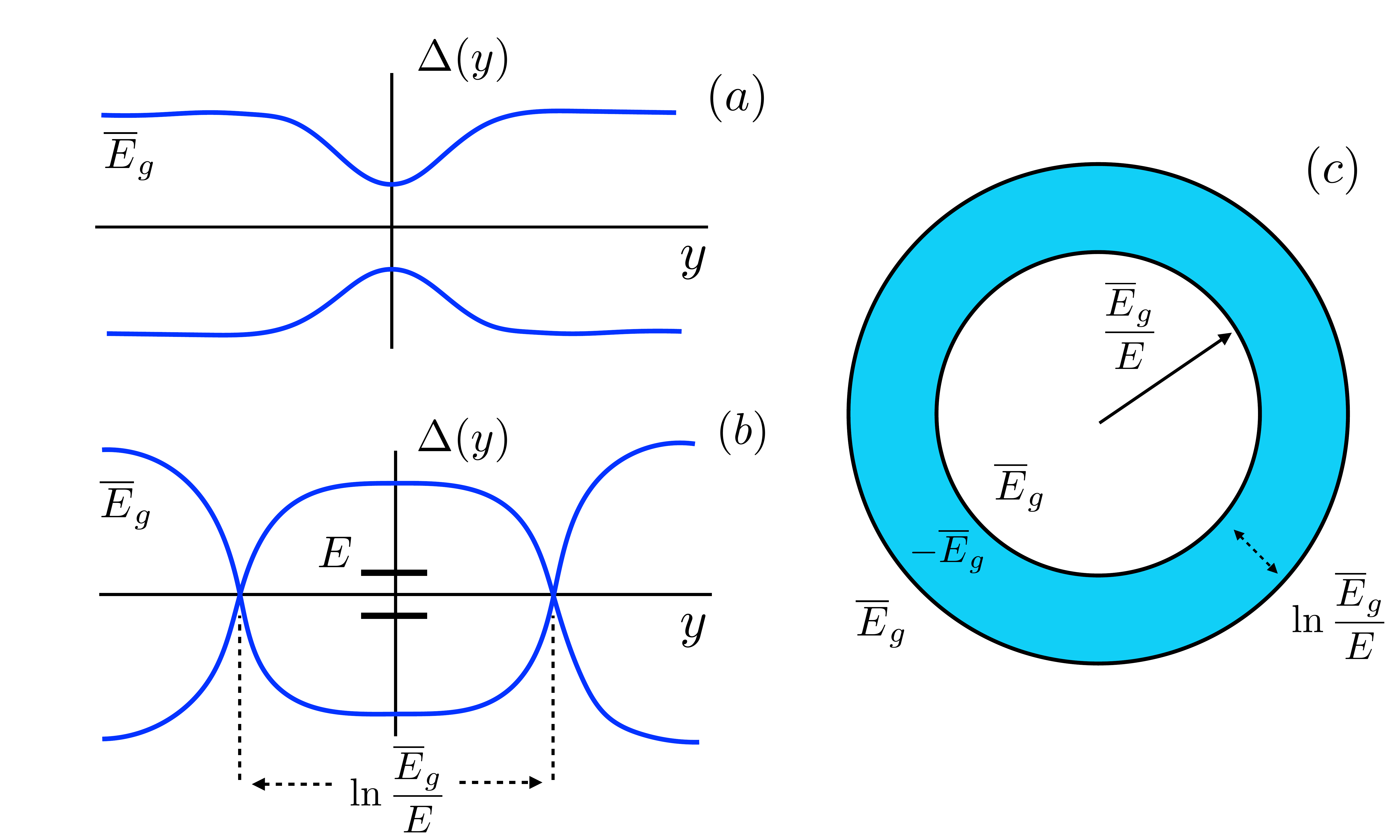}
\caption{(Color online)  (a) Fluctuation states near the gap edges  are due to local reductions of the width of the gap caused by the
 the composition disorder. (b) To create an in-gap state with energy, $E$, much smaller than the gap width in 1d {\em two} local gap reversals are required.(c) In 2d, the angular motion tends to shift the fluctuation levels away from the gap center. Thus, the fluctuation, responsible for the state, $E$, represents a narrow ring of gap reversals with radius $\propto {\overline E}_g/E$.
}\label{figure1}
\end{figure}

\noindent{\em Introduction.}
Pairs of spin-degenerate chiral edge modes
are implicit for insulators with inverted bandstructure.\cite{KaneMele,HgTe}
The minimal model\cite{HgTe} which captures these modes is $4\times 4$ matrix
Hamiltonian acting in the basis of two spin and two orbital states.

The origin of the quantum anomalous Hall effect\cite{2008} (QAH)
is breaking of the time-reversal symmetry induced by magnetic order.
As a result, the symmetry between the two
counterpropagating modes at the sample edges
is lifted.
With a single chiral mode per edge,
the Hall conductance of the sample becomes
nonzero, and the transport resembles the conventional quantum Hall effect.
Experimental
studies\cite{1,2,3,Pioneering,Chekelsky1,UCLA1,Robust,PennState,Tokura1,UCLA2,Goldhaber,Moodera1,Moodera2,Molenkamp,Zeldov,Feng15},
on Cr-doped and V-doped layers of BiTe-based insulating compounds
confirm both the quantization of the Hall resistance and the edge transport
which accompany the buildup of
the magnetic order.

For QAH effect to be pronounced, the bulk of the sample should be
strongly insulating. On the other hand, the crossover between a trivial
and ``topological" bandstructures
takes place as the gap passes through zero. Obviously,
the smaller is the gap the easier it is washed out by the disorder.
More precisely, the disorder gives rise to in-gap states.
However, in QAH, the disorder is of a special type:
randomness in positions of magnetic ions
cause the  local fluctuations of the gap {\em width}.
For such fluctuations the energies near the
gap center remain unaffected. This is probably
the reason why robust QAH is observed in experiments of
several groups.

In the present paper we study quantitatively the smearing of the gap
due to statistical, and thus unavoidable, magnetic disorder. We find that
the states near the gap center are due to the local reversals of the gap {\em sign}
within narrow rings. By employing the instanton approach\cite{Halperin,Zittarz}
we specify the shape
of these fluctuations and the
likelihood of their occurrence, which determines the density of the in-gap
states. This density of states exhibits a semi-hard gap near zero energy.

\noindent{\em Instanton approach.}
Due to the  composition disorder, the local value of
$x$, which is the portion of magnetic ions, differs from average
\begin{equation}
\label{x}
 x({\bf r})={\overline x}+\delta x({\bf r}).
\end{equation}
Fluctuations  $\delta x({\bf r})$ are gaussian with a zero
correlation radius
\begin{equation}
\label{xx}
\langle\delta x({\bf r})\delta x({\bf r}^{\prime})\rangle
=\frac{{\overline x}(1-{\overline x})}{N_0}\delta({\bf r}-{\bf r}^{\prime}),
\end{equation}
where $N_0$ is the concentration of Bi lattice sites in which the
substitution magnetic ions reside.
It is natural to assume that the local gap fluctuations, $\Delta({\bf r})$,
are proportional to $\delta x$, i.e.
\begin{equation}
\label{Delta}
\Delta({\bf r}) =E_g({\bf r})-{\overline E}_g=\alpha\hspace{0.5mm} \delta x({\bf r}), ~~~~\alpha=\frac{d\hspace{0.3mm}  {\overline E}_g}{d \hspace{0.3mm} {\overline x}}.
\end{equation}
It follows from Eq. (\ref{Delta}) that the probability to find the fluctuation $\Delta({\bf r})$
is given by
\begin{equation}
\label{P}
{\cal P}\big\{\Delta({\bf r})\big\}\propto \exp\Big[-\frac{1}{2\gamma}\int d{\bf r}\Delta^2({\bf r})\Big].
\end{equation}
where $\gamma =\frac{\alpha^2}{N_0}{\overline x}(1-{\overline x})$.

According to the instanton approach\cite{Halperin,Zittarz}, the density of states with
energy, $E$, corresponds to the maximum of the functional ${\cal P}$ among all the fluctuations
that create a level with energy, $E$.  In application to QAH effect, the wave function, $\Psi({\bf r})$,  corresponding to this level is a two-component spinor,
\begin{equation}
\Psi({\bf r})=\begin{pmatrix} \psi_e({\bf r}) \\ \psi_h({\bf r}) \end{pmatrix},
\end{equation}
which satisfies
the Schr{\"o}dinger equation
\begin{equation}
\label{Schrodinger}
\hat{h}_{\Delta({\bf r})}\Psi=E\Psi.
\end{equation}
The Hamiltonian $\hat{h}_{\Delta({\bf r})}$ is a standard Hamiltonian of the minimal model Ref. \onlinecite{HgTe}
in which only one spin component is retained. In the conventional notations\cite{HgTe} it has the form
\begin{equation}
\label{h}
\hat{h}_{\Delta({\bf r})}=A(\hat{k}_x \sigma_x+\hat{k}_y \sigma_y)+\Big(B\hat{k}^2+{\overline E}_g+\Delta({\bf r})\Big)\sigma_z,
\end{equation}
where $\sigma_x$, $\sigma_y$, and $\sigma_z$ are the Pauli matrices acting in the pseudospin space.
Relative sign of ${\overline E}_g$ and parameter $B$
determines
whether or not the chiral  modes are the eigenstates of this Hamiltonian in the presence of an edge\cite{2008}.

The procedure of minimization of the functional Eq. (\ref{P}) with restriction Eq. (\ref{Schrodinger})
is conventionally carried out\cite{Halperin,Zittarz} by introducing the Lagrange multiplier, $\lambda$, and searching for a minimum
of the auxiliary functional
\begin{equation}
\label{auxiliary}
\lambda\langle\Psi|\hat{h}_{\Delta({\bf r})}|\Psi\rangle +\frac{1}{2\gamma}\int d{\bf r}\Delta^2({\bf r}).
\end{equation}
with respect to $\Delta({\bf r})$. The minimization yields
\begin{equation}
\label{optimal}
\Delta({\bf r})=-\lambda \gamma \Big(|\psi_e({\bf r})|^2-|\psi_h({\bf r})|^2\Big).
\end{equation}
The remaining task is to substitute Eq. (\ref{optimal}) into the Schr{\"o}dinger equation,
find $\psi_e({\bf r})$,  $\psi_h({\bf r})$, substitute them into Eq. (\ref{optimal}), and evaluate ${\cal P}$
with extremal $\Delta ({\bf r})$ defined  by Eq. (\ref{optimal}).

\noindent{\em Asymptotic solution of the instanton equation.}
Assuming the azimuthal symmetry of $\Delta({\bf r})$, the solutions of Eq. (\ref{Schrodinger}) can be classified
according to the angular momentum: $\psi_e({\bf r})=i\psi_e^m(\rho)\exp(im\phi)$,
$\psi_h({\bf r})=\psi_h^m(\rho)\exp [i(m+1)\phi]$, where $\rho$ and $\phi$ are the radius and the azimuthal angle, respectively. Then the system of equations for $\psi_e^m(\rho)$ and $\psi_h^m(\rho)$ reads
\begin{align}
\label{System}
&\Big[{\overline E}_g-E-\lambda \gamma\Big(|\psi_e^m(\rho)|^2-|\psi_h^m(\rho)|^2\Big)\Big]\psi_e^m(\rho)
\nonumber \\
&~~~~=A\Big(\frac{\partial  }{\partial \rho}+\frac{m+1}{\rho}\Big)\psi_h^m(\rho), \nonumber \\
&\Big[{\overline E}_g+E-\lambda \gamma\Big(|\psi_e^m(\rho)|^2-|\psi_h^m(\rho)|^2\Big)\Big]\psi_h^m(\rho)
\nonumber \\
&~~~~=A\Big(\frac{\partial  }{\partial \rho}-\frac{m}{\rho}\Big)\psi_e^m(\rho).
\end{align}
Here we dropped the term $B{\hat k}^2$ in the Hamiltonian ${\hat h}_{\Delta({\bf r})}$ and will
incorporate it later.\cite{Supplemental}

The solution of the system is straightforward when the energy, $E$, is close to the band-edge, $({\overline E}_g-E)\ll {\overline E}_g$. Then we have $\psi_h^m \ll \psi_e^m$, so that the system Eq. (\ref{System}) reduces to a single equation
\begin{equation}
\label{standard}
\frac{A^2}{2{\overline E}_g}\nabla^2 \psi_e({\bf r})+\lambda \gamma \big[\psi_e({\bf r})\big]^3=({\overline E}_g-E)\psi_e({\bf r}).
\end{equation}
This is a standard instanton equation for a particle moving in a white-noise random potential.\cite{Halperin,Zittarz,Thouless,Brezin} The radial size of this instanton is $\sim \big[{\overline E}_g({\overline E}_g-E)/A^2\big]^{1/2}$. Thus, the integral over ${\bf r}$ in Eq. (\ref{P}) is proportional to $({\overline E}_g-E)$.
The full expression for the density of states in the tail reads
\begin{equation}
\label{result1}
N(E) \propto \exp \Big[ -\kappa_0 \Big(\frac{A^2}{\gamma}\Big)\frac{{\overline E}_g-E}{{\overline E}_g}\Big],
\end{equation}
so that the characteristic tail energy is given by
$E_t=\frac{\gamma {\overline E}_g}{A^2}$.
The value of the numerical factor $\kappa_0=5.8$ was established in Refs. \onlinecite{Thouless}, \onlinecite{Brezin}. It originates from the solution of Eq. (\ref{standard}) with zero angular momentum. Physically, the tail states are due to local gap reductions, as depicted
in Fig. \ref{figure1}.

The result Eq. (\ref{result1}) applies when the tail energy is much smaller than the gap, i.e. for $\gamma \ll A^2$.
This result cannot be used even as an order-of-magnitude estimate for the middle of the gap.
This is because the shape of the fluctuation, $\Delta({\bf r})$, at $|E|\ll {\overline E}_g$ is
dramatically different from a simple gap reduction, $\Delta({\bf r})$, captured by Eq. (\ref{standard}).
Below we demonstrate that for $|E|\ll {\overline E}_g$ the expression for the density of states has the
form
\begin{equation}
\label{result2}
N(E)\propto \exp\Bigg[-\kappa_1\Big(\frac{A^2}{\gamma}\Big)\frac{{\overline E}_g}{|E|}\ln\frac{{\overline E}_g}{|E|}\Bigg].
\end{equation}
Singular energy dependence of $|\ln N(E)|$ reflects the fact that
the probability of formation of a state near the gap
center is highly unlikely since the corresponding fluctuation
requires a local gap reversal.

To create a state exactly at $E=0$ the gap should be negative in the
left half-space and positive in the right half-space\cite{Grinstein} (or vice versa).
Naturally, such a fluctuation has a zero probability. To have a finite probability,
the fluctuation must include two gap reversals, i.e. $|\Delta ({\bf r})|$ should
exceed ${\overline E}_g$ inside the fluctuation. To establish the shape of this
fluctuation, we start from a one-dimensional case when  $|\Delta ({\bf r})|$ changes
only along the coordinate, $y$, Fig. \ref{figure1}.

\begin{figure}
\includegraphics[width=90mm]{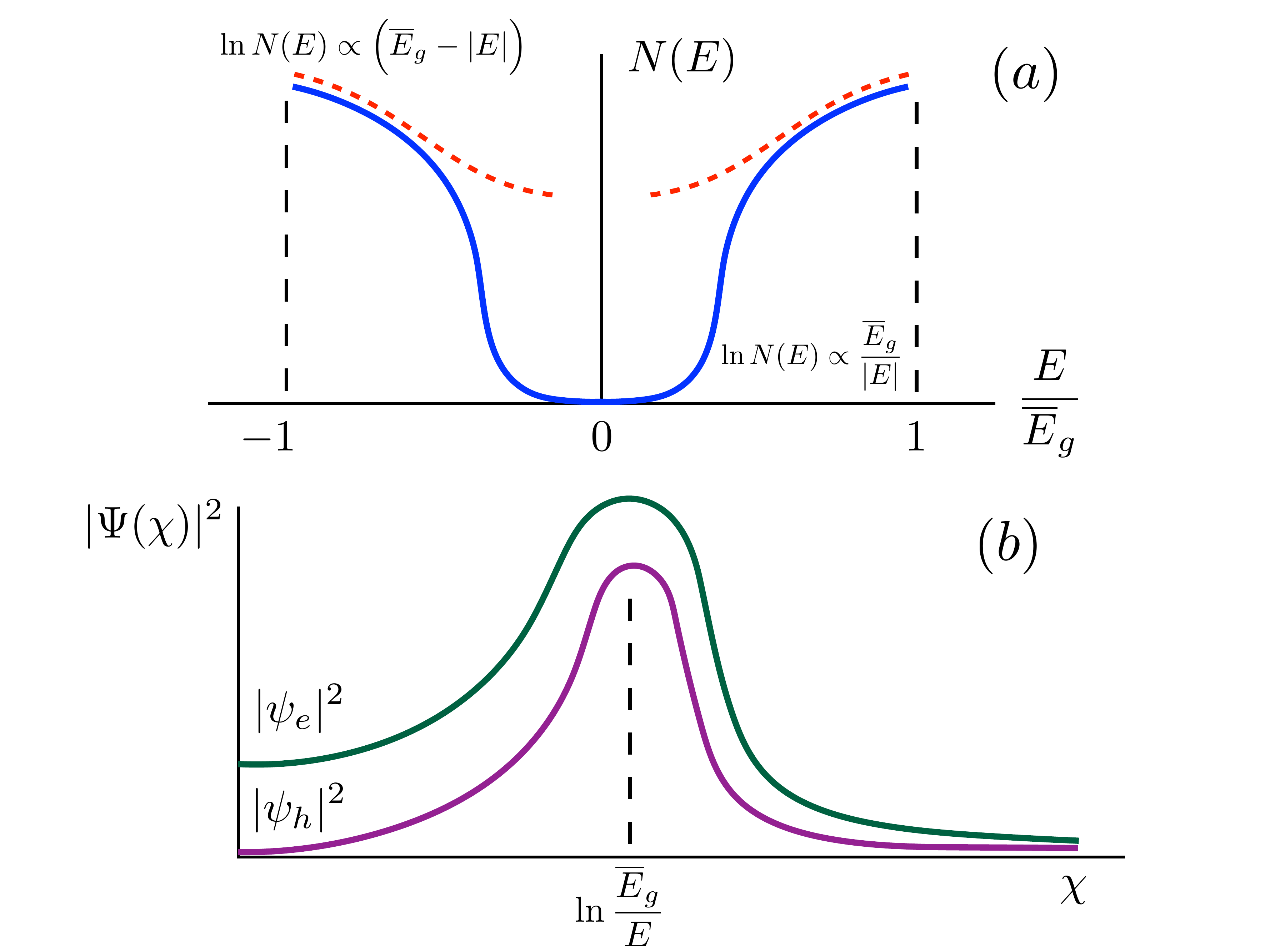}
\caption{(Color online) (a) The density of the in-gap states due to the composition disorder is shown schematically. Near the gap edges (dashed lines) it is a simple exponent, see Eq. (\ref{result1}), while near $E=0$ it represents a semi-hard gap, Eq. (\ref{result2}).
(b) The components, $\psi_e$ and $\psi_h$, of the spinor corresponding to the fluctuation state, $E\ll {\overline E}_g$, are shown schematically versus the dimensionless distance from the ring center. Analytically, $\psi_e$ is given by Eq. (\ref{dimensional}).
 The width of the ring exceeds logarithmically the in-gap decay length at $E=0$.}
\label{figure2}
\end{figure}

A one-dimensional version of the system Eq. (\ref{System}) reads
\begin{align}
\label{System1}
&\Big[{\overline E}_g-E-\lambda \gamma\Big(|\psi_e(y)|^2-|\psi_h(y)|^2\Big)\Big]\psi_e(y)
=A\frac{d\psi_h(y) }{d y},
 \nonumber \\
&\Big[{\overline E}_g+E-\lambda \gamma\Big(|\psi_e(y)|^2-|\psi_h(y)|^2\Big)\Big]\psi_h(y)
=A\frac{ d \psi_e(y) }{d  y}.
\end{align}
Upon a natural rescaling
\begin{equation}
y=\frac{A}{{\overline E}_g}\chi,~~~
\varepsilon=\frac{E}{{\overline E}_g},~~~
\psi_{e,h}=\Big(\frac{{\overline E}_g}{\lambda \gamma}\Big)^{1/2}\Phi_{e,h},
\end{equation}
it acquires a fully dimensionless form
\begin{align}
\label{System2}
&\Big[1-\varepsilon-\Big(|\Phi_e(\chi)|^2-|\Phi_h(\chi)|^2\Big)\Big]\Phi_e(\chi)
=\frac{d \Phi_h}{d\chi},
 \nonumber \\
&\Big[1+\varepsilon-\Big(|\Phi_e(\chi)|^2-|\Phi_h(\chi)|^2\Big)\Big]\Phi_h(\chi)
=\frac{d \Phi_e }{d \chi}.
\end{align}
Dimensionless length in Eq. (\ref{System2}) corresponds to a physical decay length of the
wave function in the middle of the gap.  Local gap reversals correspond to the regions of
$\chi$ where $\Big(|\Phi_e(\chi)|^2-|\Phi_h(\chi)|^2\Big)$ exceed $1$. A formal reason why
there are no midgap fluctuation states is that for $\hspace{0.2mm}$ $\varepsilon=0$ electron-hole symmetry requires $|\Phi_e(\chi)|^2=|\Phi_h(\chi)|^2$, which is incompatible with decay of $\Phi_{e,h}$
at $\chi \rightarrow \pm \infty$.

To find an asymptotic solution of the system at finite energy,
we make use of the smallness of parameter, $\varepsilon$. As a first step, instead of $\Phi_e$ and $\Phi_h$, we introduce new functions
\begin{equation}
\Phi_e(\chi)=C(\chi)\cosh \varphi(\chi),~~\Phi_h(\chi)=-C(\chi)\sinh \varphi(\chi),
\end{equation}
after which the system takes the form
\begin{align}
\label{System3}
&1-\varepsilon-C(\chi)^2=-\bigg[\frac{d \varphi}{d \chi}+\tanh\varphi(\chi)\Big(\frac{d C}{C d\chi}\Big)\bigg],
 \nonumber \\
&1+\varepsilon-C(\chi)^2=-\bigg[\frac{d \varphi}{d \chi}+\frac{1}{\tanh\varphi(\chi)}\Big(\frac{d C}{C d\chi}\Big)\bigg].
\end{align}
Upon subtracting the two equations, we can express the function $C(\chi)$ in terms of $\varphi(\chi)$
as follows
\begin{equation}
\label{C}
C(\chi)=C_0\exp\Big[-\varepsilon \int\limits_0^\chi d\chi' \sinh2\varphi(\chi')\Big],
\end{equation}
where $C_0$ is a constant. Substituting this expression back into the system, we arrive to a closed differential-integral equation
for $\varphi(\chi)$
\begin{equation}
\label{varphi}
1-C_0^2\exp\Big[-2\varepsilon \int\limits_0^\chi d\chi' \sinh2\varphi(\chi')\Big]=-\frac{d\varphi}{d\chi}+\varepsilon \cosh2\varphi.
\end{equation}
In a zero order in $\varepsilon \ll 1$ the solution of Eq. (\ref{varphi}) is a linear function
\begin{equation}
\label{zero}
\varphi(\chi)=\frac{\chi}{b},~~~~~~~b=\frac{1}{C_0^2-1}.
\end{equation}
Gap reversal, which corresponds to $C_0 >1$, is terminated at certain distance $\chi=\chi_{\varepsilon}$
when $\varepsilon\cosh \varphi$ becomes big. This yields
$\chi_{\varepsilon}=\frac{b}{2}|\ln\varepsilon|$. Importantly, the exponential  term in
the left-hand side of Eq. (\ref{varphi}) drops abruptly from $1$ to $0$ at the same $\chi=\chi_{\varepsilon}$,
or, more precisely, in the domain $|\chi-\chi_{\varepsilon}|\lesssim 1$.

The behavior of $C(\chi)$ at $|\chi-\chi_{\varepsilon}|> 1$ can be found taking into account that
$C_0^2$-term in Eq. (\ref{varphi}) is negligible in this domain. Then it follows from Eq. (\ref{varphi}) that the function $\varphi(\chi)$ saturates at the value $\varphi=\varphi_{\varepsilon}$, such that
$\varepsilon\cosh 2\varphi_{\varepsilon}=1$. Smallness of $\varepsilon$ allows to simplify $\varphi_{\varepsilon}$ to $\frac{1}{2}|\ln \varepsilon|$. This is exactly the same value which one
obtains upon substituting $\chi_{\varepsilon}$ into Eq. (\ref{zero}). The fact that $\varphi(\chi)$ saturates at $\chi >\chi_{\varepsilon}$ suggests that the function $C(\chi)$ falls off exponentially,
as $\exp\big[-(\chi-\chi_{\varepsilon})\big]$ at $\chi >\chi_{\varepsilon}$, as follows from Eq. (\ref{C}). The behavior of $\varphi(\chi)$ and $C(\chi)$ is depicted in Fig. \ref{figure2}.

Returning to dimensional units, we summarize our result
\begin{equation}
\label{dimensional}
\Phi_e(y)=\Bigg[\frac{{\overline E}_g}{\lambda \gamma}\Big(1+\frac{1}{b}\Big)\Bigg]^{1/2}\hspace{-3mm}\times
\begin{cases}
 \cosh\Big(\frac{{\overline E}_g y}{Ab}\Big),~~ y<y_{\s E}, \\
\Big( \frac{{\overline E}_g}{E}\Big)^{1/2}\hspace{-3mm}\exp\hspace{-0.5mm}\big[\hspace{-1mm}-\hspace{-0.5mm}\frac{{\overline E}_g}{A}(y-y_{\s E})\big],~y>y_{\s E}.
 \end{cases}
\end{equation}
where $y_{\s E}= \frac{A}{{\overline E}_g}\chi_{\varepsilon}=\frac{Ab}{2{\overline E}_g}\ln\frac{{\overline E}_g}{E}$. The corresponding expression for $\psi_{h}$ differs by replacement of $\cosh$
by $\sinh$. At $y=y_{\s E}$ the two expressions match within a numerical factor.

The solution Eq. (\ref{dimensional}) of the system of instanton equations in 1d is, actually, sufficient to find the
2d density of states. Compared to the system Eq. (\ref{System1}), the 2d instanton equations contain the extra ``centrifugal" terms $\propto 1/\rho$. These terms create an energy shift $\sim A/\rho$, and thus prevent the
formation of the fluctuation in-gap levels with small energies.
For such levels to exist the double reversal of the gap sign should take place within a ring with radius,
$\rho_{\s E}$, much bigger than the width, see Fig.~\ref{figure1}. Then the solutions of the system Eq. (\ref{System1}) near the
ring center are {\em one-dimensional} with $y=\rho-\rho_{\s E}$. More quantitatively\cite{Supplemental}, with angular motion taken into account, the
energy levels of the ring-shape fluctuation are given by
\begin{equation}
\label{radial}
E_m=\pm \Bigg[A^2\Big(\frac{2m+1}{\rho_{\s E}}\Big)^2+E^2\Bigg]^{1/2},
\end{equation}
where the first term described the quantization of the angular kinetic energy. The above equation suggests that, for
level $E$ to exist, the minimal radius of the ring is $A/|E|$.

In the expression Eq. (\ref{P}) for the density of states the integral $d{\bf r}$ can be replaced by the integral $2\pi\rho_{\s E}dy$ over the area of the ring
\begin{equation}
\label{N1}
N(E)\propto \exp\Bigg[-\frac{1}{2\gamma}\Big(2\pi\rho_{\s E}\int\limits_{-\infty}^{\infty}dy \Delta^2(y)\Big)\Bigg].
\end{equation}
The expression for magnitude of the fluctuation, $\Delta(y)$, is given by Eq. (\ref{optimal}).
Taking into account that the dominant contribution to the integral over $y$ comes from the domain
$|y|<y_{\s E}$, we get
\begin{equation}
\label{N2}
N(E)\propto \exp\Bigg[-2\pi\lambda^2\gamma\rho_{\s E}\int\limits_0^{y_{\s E}}dy\Big(\psi_e^2(y)-\psi_h^2(y)\Big)^2\Bigg].
\end{equation}
Substituting  Eq. (\ref{dimensional}) into Eq. (\ref{N1}), and taking into account that the difference $\left(\psi_e^2-\psi_h^2\right)$ is constant for $y<y_{\s E}$, we reproduce the result Eq. (\ref{result2}) in which the constant $\kappa_1$ should be identified with a combination

\begin{equation}
\kappa_1(b)=\pi\Big(1+\frac{1}{b}\Big)^2b.
\end{equation}
The dependence $\kappa_1(b)$ has a minimum at $b=1$, where it is equal to $4\pi$. The value $b=1$
corresponds to $C_0=2^{1/2}$, which, in turn, means that the most probable fluctuation corresponds
to a complete gap reversal, i.e. the gap is equal to $-{\overline E}_g$ at the core of the fluctuation.

\noindent{\em Discussion.}
The main outcome of our study is that, even when the spins of magnetic dopants are fully aligned, the
unavoidable statistical fluctuations in their density (alloy disorder\cite{Alloy1,Alloy2,Alloy2a,Alloy3})
smear the gap, ${\overline E}_g$. In conventional semiconductor mixed crystals this disorder is known to cause the
tail in the optical absorption and even turn a gapless semiconductor into a metal\cite{Alloy2a}.  The degree
of smearing is governed by a dimensionless material parameter
\begin{equation}
\label{nu}
\nu=\frac{\gamma}{A^2}=\frac{{\overline x}(1-{\overline x})}{A^2N_0}\Big(\frac{d{\overline E}_g}{d{\overline x}}\Big)^2.
\end{equation}
For $\nu \ll 1$ only a narrow energy interval $|E-{\overline E}_g|\sim \nu {\overline E}_g$ is affected
by the disorder, see Eq. (\ref{result1}). As $\nu$ exceeds $1$, it might seem from Eq. (\ref{result1}) that the gap is completely washed out. However, our result Eq. (\ref{result2}), see also Fig. \ref{figure2}, suggests that, even for strong disorder, there is an almost hard gap near $E=0$ which exists in the domain
$|E|\lesssim {\overline E}_g/\nu$. Probably,\cite{Supplemental} it is this hard gap that governs the temperature dependence of the longitudinal resistance in the experiments.
\cite{1,2,3,Pioneering,Chekelsky1,UCLA1,Robust,PennState,Tokura1,UCLA2,Goldhaber,Moodera1,Moodera2,Molenkamp,Zeldov,Feng15}
The scale of temperatures for QAH effect is known to be much lower than the Curie temperature.
The fact that the bulk gap in QAH is narrow follows most convincingly from Ref. \onlinecite{UCLA1}, where the strong
temperature-dependent deviations from the quantized value of non-diagonal resistance were observed in high applied external field, so that they cannot be accounted for by the  domain structure in the sample.
Moreover, the analysis in experimental paper Ref. \onlinecite{Goldhaber} indicates that the low-temperature behavior of the zero-field diagonal conductivity
is activational rather than the variable-range hopping, which is consistent with the scenario of a hard gap.
To estimate the experimental value of parameter $\nu$ we chose $x=0.1$, as in most experiments, $A=3$eV${\AA}$
(Ref. \onlinecite{2008}), and $\alpha=2.7$eV (Ref. \onlinecite{Numbers}). With $N_0=5\cdot 10^{14}$cm$^{-2}$, we got
$\nu \approx 0.5$, suggesting that  statistical disorder is relevant for QAH effect.

There are two principle issues that complicate quantitative comparison of our predictions with experiment.
Firstly, we used the simplest description of electron states based on the Hamiltonian Eq. (\ref{h}).
This Hamiltonian, proposed in Ref. \onlinecite{Numbers}, is believed to capture the low-energy excitations
 after the pseudospin components, $\psi_e$ and $\psi_h$, are identified with symmetric and antisymmetric combinations
of the top and bottom surface states.  However, the experiments were performed on multilayer structures. It is unclear
whether the purely 2d description applies to them quantitatively. Secondly, in realistic samples, the in-gap states due to the magnetic  disorder can be masked by the smearing  due to non-magnetic impurities. In-gap states due to these impurities do not ``preserve"
 the energy $E=0$.  The only information about the disorder in
QAH samples is the value of mobility, $\mu =760$cm$^2/$Vs, measured in Ref. \onlinecite{Pioneering} at temperature $80$K, much higher than
the Curie temperature, $15$K. However, relating this mobility to the random potential, which could be added to the Hamiltonian Eq. (\ref{h}), is impossible, again, due to the complex bandstructure of multilayers.

In conclusion, we point out that for a really strong disorder $\nu \gg 1$, the hard gap near $E=0$ disappears. In this limit one can neglect ${\overline E}_g$ in the Hamiltonian, so that the problem reduced to disorder induced smearing
of a linear Dirac spectrum. This problem has a long history,\cite{FisherFradkin,PALee1993,TsvelikWenger}
 and was addressed in relation to e.g. $d$-wave superconductivity. However, in the absence of energy scale to compare
 the disorder with, there is no definite answer.

\noindent{\em Acknowledgements.}
We are grateful to D. A. Pesin for an illuminating discussion on the relevance of the Hamiltonian Eq. (\ref{h}). We are also grateful to
Jing Wang (Stanford University) for introducing  the topic  of  QAH effect to us.  This work was supported by NSF through MRSEC DMR-1121252.

\newpage
\section{Supplemental material}

\subsection{Quantized levels on a ring with inverted bandgap}
Consider a gap-inverting fluctuation  confined to a ring $\rho_1 <\rho <\rho_2$.
More specifically, the gap, $\Delta({\bf r})$, changes in a radial direction as follows:
\begin{equation}
\Delta({\bf r})=
\begin{cases}
 \Delta_0,~~ ~~~0<\rho<\rho_1, \\
-\Delta_0,~~~\rho_1<\rho<\rho_2,\\
\Delta_0,~~ ~~~\rho>\rho_2.
 \end{cases}
\end{equation}
We assume for simplicity that the gap reversal is full. In the domain $\rho<\rho_1$, the in-gap solution of the system Eq. [10] of the main text which is finite at the
origin reads
\begin{equation}
\begin{pmatrix} \psi_e^m(\rho) \\ \psi_h^m(\rho) \end{pmatrix}=
\begin{pmatrix} \alpha I_{m+1}(s\rho) \\ \sqrt{\frac{\Delta_0-E}{\Delta_0+E}}\alpha I_m(s\rho) \end{pmatrix},
\end{equation}
where $I_m(z)$ is the modified Bessel function, and $s=\frac{\sqrt{\Delta_0^2-E^2}}{A}$. Corresponding solution for $\rho >\rho_2$, which decays at $\rho\rightarrow \infty$ can expressed in terms of the Macdonald function
as follows
\begin{equation}
\begin{pmatrix} \psi_e^m(\rho) \\ \psi_h^m(\rho) \end{pmatrix}=
\begin{pmatrix} \beta K_{m+1}(s\rho) \\ -\sqrt{\frac{\Delta_0-E}{\Delta_0+E}}\beta K_m(s\rho) \end{pmatrix}.
\end{equation}
Within the ring, the solution is a linear combination of $I_m(s\rho)$ and $K_m(s\rho)$
\begin{equation}
\begin{pmatrix} \psi_e^m(\rho) \\ \psi_h^m(\rho) \end{pmatrix}=
\begin{pmatrix} \alpha_1 I_{m+1}(s\rho)+\beta_1K_{m+1}(s\rho) \\ \sqrt{\frac{\Delta_0+E}{\Delta_0-E}}\Big(-\alpha_1 I_m(s\rho) +\beta_1K_m(s\rho)\Big) \end{pmatrix},
\end{equation}
The gap inversion is reflected in the relative signs  of the components of the spinor inside and outside the ring.
Four unknown coefficients, $\alpha$, $\beta$, $\alpha_1$, and $\beta_1$ are related by continuity of the components
of the spinors at $\rho=\rho_1$ and $\rho=\rho_2$. Energy levels are determined from the condition of consistency of the system,
which reads
\begin{widetext}
\begin{align}
\label{levels}
\Big(\frac{\Delta_0-E}{\Delta_0+E}\Big)\frac{K_{m}(s\rho_2)}{K_{m+1}(s\rho_2)}=
\frac{\Big(I_{m+1}(s\rho_1)K_m(s\rho_1)-\frac{\Delta_0-E}{\Delta_0+E}I_m(s\rho_1)K_{m+1}(s\rho_1)\Big)I_m(s\rho_2)-
\frac{2\Delta_0}{\Delta_0+E}I_{m+1}(s\rho_1)I_m(s\rho_1)K_m(s\rho_2)}
{\Big(I_{m+1}(s\rho_1)K_m(s\rho_1)-\frac{\Delta_0-E}{\Delta_0+E}I_m(s\rho_1)K_{m+1}(s\rho_1)\Big)I_{m+1}(s\rho_2)+
\frac{2\Delta_0}{\Delta_0+E}I_{m+1}(s\rho_1)I_m(s\rho_1)K_{m+1}(s\rho_2)}.
\end{align}
\end{widetext}
The near-midgap levels with $|E|\ll \Delta_0$ appear only when the conditions $s\rho_1\gg 1$ and $s(\rho_2-\rho_1)\gg 1$ are met.
Under these conditions Eq. (\ref{levels}) allows serious simplifications. Firstly, using the asymptotes of the Bessel functions,
the common bracket in the numerator and denominator of the right-hand side simplifies to
\begin{align}
\label{bracket}
I_{m+1}(s\rho_1)K_m(s\rho_1)&-\Big(\frac{\Delta_0-E}{\Delta_0+E}\Big)I_m(s\rho_1)K_{m+1}(s\rho_1)\nonumber \\
&\approx\frac{1}{2s\rho_1}\Big(\frac{2E}{\Delta_0}-\frac{2m+1}{s\rho_1}\Big).
\end{align}
As a next step, we divide both sides by the ratio $I_m(s\rho_2)/I_{m+1}(s\rho_2)$ and take the large-$\rho$ asymtotes. Then the
left-hand side takes the form
\begin{equation}
\label{left}
\Big(\frac{\Delta_0-E}{\Delta_0+E}\Big)\frac{K_{m}(s\rho_2)I_{m+1}(s\rho_2)}{K_{m+1}(s\rho_2)I_{m}(s\rho_2)}=
1-\frac{2E}{\Delta_0}-\frac{2m+1}{s\rho_2}.
\end{equation}
The expressions in the numerator and denominator of the right-hand side are equal to Eq. (\ref{bracket}) $\pm$ a small correction. The asymptotic form
of this correction is the following
\begin{align}
\label{correction}
&\frac{2\Delta_0}{\Delta_0+E}I_{m+1}(s\rho_1)I_m(s\rho_1)\frac{K_m(s\rho_2)}{I_m(s\rho_2)}
\nonumber \\
&\approx\frac{2\Delta_0}{\Delta_0+E}I_{m+1}(s\rho_1)I_m(s\rho_1)\frac{K_{m+1}(s\rho_2)}{I_{m+1}(s\rho_2)}\approx
\frac{2e^{-2s(\rho_2-\rho_1)}}{2s\rho_1}.
\end{align}
Upon combining Eqs. (\ref{bracket})-(\ref{correction}), the equation for the energy levels reduces to
\begin{equation}
1-\frac{2E}{\Delta_0}-\frac{2m+1}{s\rho_2}=1-\frac{4\exp\big[-2s(\rho_2-\rho_1)\big]}{\frac{2E}{\Delta_0}-\frac{2m+1}{s\rho_1}}.
\end{equation}
For a narrow ring one can replace  $\rho_1$ and $\rho_2$ in the denominators by $(\rho_1+\rho_2)/2$. Also, with accuracy $E^2/\Delta_0^2$,   one can replace $s$ by $\Delta_0/A$. This leads to the following expression for the energy levels
\begin{equation}
\Bigg(\frac{2E}{\Delta_0}\Bigg)^2=\Bigg[\frac{2(2m+1)A}{\Delta_0(\rho_1+\rho_2)}\Bigg]^2+4\exp\Big[-2\frac{(\rho_2-\rho_1)\Delta_0}{A}\Big].
\end{equation}
The right-hand side is the sum of contributions from the quantized motion along the ring and quantized motion across the ring, as in
Eq. [23] of the main text.

\subsection{Temperature dependence of conductivity }
Neglecting the energy dependence of logarithm in Eq. [13] of the main text, we approximate the energy-dependent  density of states with
\begin{equation}
\label{N}
N(E)=N_0\exp\Big(-\frac{{\overline E}_g}{\nu E}\Big),
\end{equation}
where parameter $\nu$ is defined by Eq. [27] of the main text.
Assume that the energy responsible for the transport is $E_0$. The density of states can be treated as a constant within a strip
$|E-E_0|<\nu E_0^2/{\overline E}_g$. A typical distance between the localized states within the strip is
\begin{equation}
r(E_0) =  \Big(\frac{\nu E_0^2N(E_0)}{{\overline E}_g}\Big)^{-1/2}.
\end{equation}
 Following
the derivation of  Mott's law, we minimize the log-resistance,
\begin{equation}
\label{conductivity}
\ln R\left(E_0\right)= \frac{E_0}{T}+\frac{2r(E_0)}{\xi},
\end{equation}
corresponding to activation into the strip and
tunneling between the neighbors, with respect to $E_0$.
Here $\xi$ is the localization radius. The condition of minimum reads
\begin{equation}
\label{conductivity1}
\frac{1}{T}=\frac{1}{\left(N_0\xi^2\right)^{1/2}}\Big(\frac{{\overline E}_g}{\nu E_0^2}\Big)^{3/2}\exp\left(\frac{{\overline E}_g}{2\nu E_0}\right),
\end{equation}
where we have differentiated only the exponent in $r(E_0)$.
Upon expressing $E_0$ from Eq. (\ref{conductivity1}) and substituting it back into Eq. (\ref{conductivity}),
we find with logarithmic accuracy
\begin{equation}
\label{conductivitu3}
\ln R\left(E_0\right)=\frac{{\overline E}_g}{2 \nu T}\bigg[\ln\frac{{\overline E}_g}{\nu T}
\Big(\frac{N_0\xi^2{\overline E}_g}{\nu}\Big)^{1/2}\bigg]^{-1}.
\end{equation}
The result Eq. (\ref{conductivitu3}) applies
when the logarithm is big. By virtue of the same condition the activation term in Eq. (\ref{conductivity1}) exceeds the
tunneling term. Concerning the dimensionless combination, $N_0\xi^2{\overline E}_g$, under the logarithm, with localization
length, $\xi=A/{\overline E}_g$, in the middle of the gap being disorder independent, this combination is some unknown power of $\nu$. Thus, for $\nu \sim 1$, Eq. (\ref{conductivitu3}) applies for $T < {\overline E}_g$. We conclude that, due to a rapid growth
of the density of states away from the gap center, the behavior of the resistance remains Arrhenius even at low temperatures.
This is consistent with observation in Ref. [15] of the main text.

\subsection{The role of $Bk^2$ term in the Hamiltonian}
In order to estimate the effect of $Bk^2$ term in the Hamiltonian Eq. [7] of the main text, we compare it to the linear term and find the
scale of momenta $k\sim A/B$ when this term becomes important. In other words, the term $Bk^2$ plays a dominant role when the
spatial scales in the problem are $\sim B/A$. On the other hand, with logarithmic accuracy, the size of the fluctuation is $\sim A/{\overline E}_g$. The ratio of the two scales yields a dimensionless parameter $B{\overline E}_g/A^2$, which becomes progressively small as the gap
decreases. More quantitative information about the role of $Bk^2$ can be obtained upon incorporating it into Eq. [20] of the main text. Then
it takes the form
\begin{align}
\frac{d\varphi}{d\chi}&=C_0^2\exp\Big[-2\varepsilon \int\limits_0^\chi d\chi' \sinh2\varphi(\chi')\Big]-1+\varepsilon \cosh2\varphi\nonumber \\
&+\frac{B{\overline E}_g}{A^2}\Bigg[\Big(\frac{d\varphi}{d\chi}\Big)^2-\frac{d^2\varphi}{d\chi^2}\Bigg].
\end{align}
Similar to the steps in the main text, we neglect small terms containing $\varepsilon$ and substitute $\varphi (\chi)=\chi/b$. This leads
to the following modified relation between the parameters $b$ and $C_0$
\begin{equation}
\label{modified}
C_0^2=1+\frac{1}{b}-\Bigg(\frac{B{\overline E}_g}{A^2}\Bigg)\frac{1}{b^2}.
\end{equation}
Note that for the ``topological" bandstructure, when the signs of $B$ and ${\overline E}_g$ are opposite, the last term in
Eq. (\ref{modified}) causes only a slight increase of $C_0$, which results in a small suppression of the exponent in the
density of states. On the contrary, for a ``trivial" bandstructure, this last term decreases $C_0$,  leading to the enhancement of the density of states. Moreover, Eq. (\ref{modified}) suggests that this enhancement can be parametrically big when the second and the third
terms closely compensate each other. For such a compensation the width of the ring should be of the order of the minimal length $B/A$.
This, however, violates our basic assumption that the shape of the fluctuation is dominated by the inner part.

\end{document}